\documentclass[onecolumn,superscriptaddress]{revtex4}

\usepackage{graphicx,float}
\usepackage{epsfig}
\usepackage{dcolumn}
\usepackage{ulem}
\usepackage{bm}
\usepackage{amsmath}
\usepackage{amssymb}
\usepackage{hyperref}
\hypersetup{colorlinks=true,linkcolor=blue,citecolor=blue,urlcolor=black}

\begin{document}

\title{Quantum information processing with space-division multiplexing optical fibres}
\date{\today}

\author{Guilherme~B.~Xavier}
\email{guilherme.b.xavier@liu.se}
\affiliation{Institutionen f\"or Systemteknik, Link\"opings Universitet, 581 83 Link\"oping, Sweden}

\author{Gustavo~Lima}
\email{glima@udec.cl}
\affiliation{Departamento de F\'{\i}sica, Universidad de Concepci\'on, 160-C Concepci\'on, Chile}
\affiliation{Millennium Institute for Research in Optics, Universidad de Concepci\'on, 160-C Concepci\'on, Chile}

\begin{abstract}
The optical fibre is an essential tool for our communication infrastructure since it is the main transmission channel for optical communications. The latest major advance in optical fibre technology is spatial division multiplexing (SDM), where new fibre designs and components establish multiple co-existing data channels based on light propagation over distinct transverse optical modes. Simultaneously, there have been many recent developments in the field of quantum information processing (QIP), with novel protocols and devices in areas such as computing, communication and metrology. Here, we review recent works implementing QIP protocols with SDM optical fibres, and discuss new possibilities for manipulating quantum systems based on this technology.
\end{abstract}

\maketitle


Quantum information processing (QIP) is a field that has seen tremendous growth over the many years since Richard Feynman's seminal talk on the use of quantum computers to simulate physical systems \cite{Feynman}. When information bits are encoded on individual or entangled quantum states, a gain over traditional systems can be seen for some information processing tasks. A famous example is the well-known Shor's algorithm for prime number factorisation running on a quantum computer, where an impressive reduction in resources is obtained when compared to classical algorithms \cite{Shor}. Another major application of QIP is in communication security, where the fact that unknown quantum states cannot be faithfully cloned \cite{nocloning} is exploited to detect the presence of an eavesdropper. This concept was used as the core foundation behind quantum key distribution (QKD), a communication protocol designed to distribute random private keys among remote parties \cite{BB84, Gisin_2002}. As the first application to showcase in practice the benefits of QIP, QKD has experienced huge development ever since \cite{Lo_2014, Diamanti_2016, Xu_2019}. QKD is part of a more general family of protocols called quantum communications (QC), which includes other schemes such as quantum teleportation and entanglement swapping \cite{Gisin_2007}, aiming to be the communication backbone supporting future networks of quantum computers \cite{Wehner_2018}.

During the last decades a number of technological features were developed by the telecommunication community in order to support the continuous increase in demand for more transmission bandwidth over a communication channel. These developments have been motivated by several applications that have cropped up along the years, from the internet to social networking and high-quality on-demand video streaming. Arguably the optical fibre has played a major role in the success of the telecommunication infrastructure, mainly due to its high transparency and high-bandwidth support \cite{Agrawal}. Technologies such as wavelength division multiplexing (WDM) \cite{WDM}, and the erbium doped fibre amplifier (EDFA) \cite{Payne, Desurvire} have been major catalysts to the extremely high capacities and ultra-long transmission distances available today. The latest technological drive towards maintaining the bandwidth growth is called spatial division multiplexing (SDM), and it consists of employing the transverse spatial properties of a light beam to multiplex information and increase the data capacity \cite{Richardson_2013}. SDM nowadays routinely allows hundreds of Tbit/s of transmission capacity \cite{Wakayama_2019}, ensuring that the bandwidth demand can keep growing.

In photonic QIP, different degrees-of-freedom (DOFs) of a single-photon such as polarisation, frequency and its transverse momentum, are used for encoding quantum systems of arbitrary dimension \cite{Sciarrino_2018, Erhard_2018}. Of particular interest, for instance, is the strategy that relies on encoding a quantum system in terms of the transverse optical modes available, which provides versatility to define high-dimensional Hilbert spaces  \cite{Rarity_1991, Neves_2005, Boyd_2005, Groblacher_2006, Rossi_2009, Aguilar_2018}. It has already been proven to be useful for QIP with photonic integrated circuits \cite{Duan_2001, Politi_2008, Obrien_2018, Sciarrino_2018, Chen_2018}, aimed at improving the robustness and compactness of experimental setups. However, remarkably, propagation over optical fibres for such quantum states has been a major challenge due to the fact that many optical fibres are designed to support only the fundamental gaussian propagation mode. Although multi-mode fibres exist (and have been available before single-mode fibres), they support hundreds of modes, requiring complex auxiliary optoelectronic systems to reconstruct the original phase wavefront following propagation \cite{Choi_2012}. Fortunately, SDM optical fibres and components can be used to cover this gap. In this topical review we go through the different SDM technologies, and cover a number of key experiments that have recently shown the advantages and new possibilities that this technology offers for QIP. We also discuss the integration of QIP into an SDM optical fibre-based network infrastructure.

\section{SDM fibres and components for optical communications}

\subsection{SDM fibres}

A typical telecommunication optical fibre is a cylindrical dielectric silica waveguide composed of a core and the surrounding cladding, possessing extremely low losses ($<$ 0.2 dB/km at 1550 nm) \cite{Agrawal}. A major advance for increasing transmission capacity was achieved when multiple data channels were multiplexed using different wavelengths over a single fibre, a technique called wavelength division multiplexing (WDM) \cite{WDM}. Although WDM, together with other technologies, has been incredibly successful in sustaining the internet growth, optical fibres are nevertheless reaching their maximum capacities \cite{Richardson_2013}. A promising avenue to continue expanding data rates over the fibre-optical infrastructure is to explore different transverse optical modes of a light beam to multiplex data channels, thus increasing the spatial data density. This technique is the spatial analogue of WDM, and aptly named spatial division multiplexing \cite{Richardson_2013}.

\begin{figure}[t]
\centering
\includegraphics[width=0.9\textwidth]{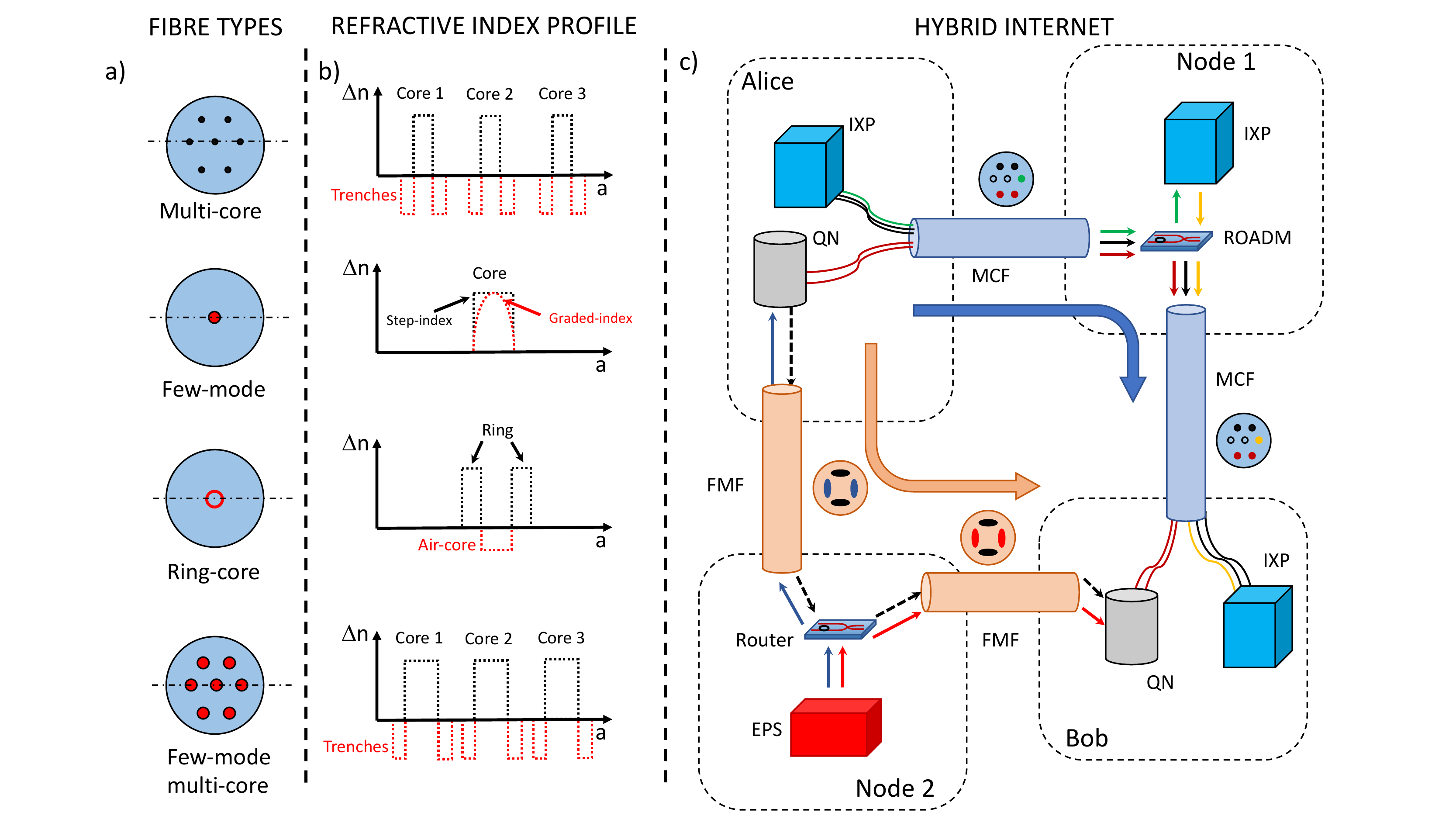}
\caption{Optical fibres for SDM-based quantum information processing. a) Cross-section schematics of four key types of SDM fibres. b) Simplified transverse relative refractive index profile $\Delta n$ as a function of radial distance $a$, taken along the dashed line in the corresponding cross-sections in a). The dashed red lines indicate geometry variations for specific fibre types. c) Hypothetical hybrid internet scheme where classical and quantum channels co-exist in SDM networks. The classical and quantum channels are allocated to different spatial modes in the fibres, i.e. different cores or transverse modes within the same core. Additional wavelength filtering is also used depending on spatial mode isolation. Alice may communicate to Bob through two possible routes, depending on current network availability. The blue route (going through intermediate node 1), employs multi-core fibres to distribute single-photons and classical traffic to Bob. Cross-sections show the different occupied cores along the links. As an example, two cores (painted red) are used as quantum channels. The other (salmon) route, uses few-mode fibres to send the single-photons through a quantum-teleportation based link, using the entangled photons produced at intermediate Node 2 as a resource. Here a router based on a photonic integrated chip is used to connect the entangled source to the two fibres, while allowing the classical information from Alice to pass through to Bob. Cross sections show the occupied LP$_{11a}$ and LP$_{11b}$ transverse modes as examples. Please see the text for details. EPS: Entangled photon source; FMF: Few-mode fibre; IXP: Internet exchange point; MCF: Multi-core fibre; QN: Quantum node; ROADM: Reconfigurable optical add-drop multiplexer.  \label{Fig1}}
\end{figure}

Any fibre that can support more than one transverse optical mode of propagation is, in principle, suitable for SDM. Different types of SDM optical fibre can be seen in Fig.~\ref{Fig1}a. The most direct approach is to embed several single-mode cores in a single fibre cladding, thus producing a multi-core fibre (MCF) \cite{Iano_1979,Saitoh_2016}. If the transverse core separation is larger than approximately 40 $\mu$m, the attenuation of cross-coupled optical power between the cores exceeds several tens of dBs, and thus they can be approximated as independent fibres in most applications. These fibres are referred to as weakly-coupled MCFs \cite{Saitoh_2016}, and the main advantage is that multiple-input multiple-output (MIMO) receivers \cite{MIMO} are not needed for detection of the spatial channels.  The price to be paid is a lower spatial channel density, since the cladding diameter cannot be much larger than approximately 200 $\mu$m to avoid rupture due to bending \cite{Saitoh_2013}. Nevertheless successful results have recently been obtained for 19-core weakly-coupled MCFs \cite{Sakaguchi_2013}.  In order to minimise inter-core coupling, lower refractive-index ``trenches'' or holes are often used around the cores (Fig. \ref{Fig1}b), at the once again expense of lower spatial density.

Another alternative is to use a fibre with a single core, but which is capable of supporting more than one transverse mode for light propagation \cite{Berdague_1982}. Standard telecommunication multi-mode optical fibres (MMFs) are actually not very useful in this regard since they support many propagation modes requiring MIMO detection systems with extremely high complexities \cite{Saitoh_2016}. Recently, significant progress has been achieved using fibres that only support a few linearly polarised (LP) transverse modes of propagation (typically 3 or 6), the so-called few-mode fibres (FMFs) \cite{Sillard_2014}. The main difference with an SMF is a larger core area, and borrowing from MMF technology, parabolic refractive index profiles can be used to minimise mode group velocity dispersion (Fig. \ref{Fig1}b). With only a few modes present intermodal crosstalk is limited and thus MIMO decoding is feasible. Significant progress has been made recently, combining MCFs with FMFs technology, yielding fibres with multiple cores where each core can support a few modes (FM-MCFs, see Fig. \ref{Fig1}a) \cite{Xia_2012}. Due to an increased spatial channel density, these fibres can greatly increase the transmission capacity \cite{van_Uden_2014}. Finally, owing to their multi-core nature, and due to the fact that the higher-order modes have a larger mode field diameter (MFD) than the fundamental LP$_{01}$ mode, lower refractive index trenches are also employed to minimise cross-talk from adjacent cores \cite{Xia_2012, van_Uden_2014}.

Another strategy for spatial multiplexing of data channels has been the use of transverse optical modes carrying orbital angular momentum (OAM) \cite{Gibson_2004}. A Laguerre-Gaussian beam carries discrete orbital orthogonal angular momentum modes characterised by an integer $l$, called the topological charge. Each associated photon carries an OAM of $l\hbar$, where $\hbar$ is Planck's constant divided by $2\pi$ \cite{Allen_1992}. Recently, OAM-encoded beams have been used for spatial multiplexing and to demonstrate the possibility of Tbit/s data rates over free-space links \cite{Wang_2012}. Following that, demonstrations of the transmission of these beams over optical fibres with high-index ring refractive index profiles were reported \cite{Bozinovic_2013}, thus expanding the toolbox of such optical modes towards fibre communications. These fibres are usually referred to as ring- or air-core fibres \cite{Brunet_2015, Gregg_2015}, due to their characteristic high refractive-index ring profile (Fig. \ref{Fig1}b).

\subsection{Multiplexers and demultiplexers}

Multiplexers and demultiplexers (also typically called fan-in and fan-out devices respectively) are employed to combine and split different data streams into corresponding spatial channels in an SDM fibre. Here we shall only focus on passive components that can already be implemented directly integrated within the fibres or through photonic chips (i.e. without resorting to bulk optical elements). In common, they all take $N$ independent single-mode input fibres, and map them onto a particular mode on the SDM fibre. For MCFs, mux/demuxes can be directly constructed using integrated waveguides written three-dimensionally onto silica chips using ultrafast laser writing \cite{Thomson_2012}. The appropriate fibres are then connected to the chip. Alternatively, discrete components (based on fibre bundles or compact lenses for example) may be used \cite{Watanabe_2012, Tottori_2012}. For FMFs, devices called photonic lanterns are normally used \cite{Birks_2015}, where the input single-mode fibres are tapered together (in parallel) ending in a multi-mode fibre at the other end. If the single-mode fibres have slightly different sized cores, then a mode-selective lantern can be built \cite{Yerolatsitis_2014}, yielding considerably better mode excitation/separation than using identical single-mode inputs/outputs. Recently, lanterns built onto integrated photonic chips have also been developed, allowing better integration possibilities \cite{Riesen_2014}. For OAM-carrying optical modes, there are demultiplexers, typically called mode sorters, and they have been done mostly using bulk optic components/active elements \cite{Berkhout_2010}. Recently, significant progress has been made towards all-fibre OAM sorters, thus making possible the use of compact and passive multiplexing elements for OAM carrying optical modes as well \cite{Zeng_2018}. Finally, fan-in/fan-out units for FM-MCF fibres can also be constructed in integrated photonics chips resorting to the 3D-femtosecond laser writing technique and integrated lanterns \cite{Riesen_2017}.

The previously mentioned SDM technologies can already be used to support a ``hybrid internet'', where quantum and classical communication systems co-exist (Fig. \ref{Fig1}c). Here, as an example, one party (Alice) wishes to send information encoded on single-photons to a remote party (Bob). Specifically they both have quantum nodes (QNs), which might be a quantum computer or a QKD terminal. The single-photons are to be transmitted using the telecommunication fibre-optical infrastructure, where classical data streams will be simultaneously transmitted to maximise the use of the available infrastructure. This is represented by IXPs (internet exchange points), where several local data streams are aggregated for transmission over a high-capacity optical link to another IXP for delivery. Two routes are possible in this example for Alice's single-photons to reach Bob: (i) the blue clockwise direction going through an intermediate node with a reconfigurable add and drop multiplexer (ROADM), where one of the classical core channels is dropped to the intermediate IXP, and a new one is added towards the IXP at Bob. The rest of the traffic (both quantum and classical) is forwarded to Bob. The counter-clockwise route (ii) is used by the QNs and it consists of a link enhanced by quantum repeaters \cite{Gisin_2007} (a simplified version is shown where a single entangled photon pair is shared across the link). Here,  one of the modes (LP$_{11a}$) is used by the entangled photons (blue from the entanglement source to Alice, and red to Bob), while the LP$_{11b}$ mode is used for the classical channel from Alice to Bob for synchronisation and communication information required by quantum repeater protocols. Additional wavelength separation may be used to avoid cross-mode contamination. The two links employ different SDM technologies, as the blue one is based on MCFs (with 2 cores used by the QNs, and 3 by the IXPs), while the salmon one is deployed with FMFs. This illustrates a possible case in network channel allocation, where routes are dynamically assigned depending on availability. For this to be feasible, the network must be transparent in terms of the employed SDM technology.

\section{Quantum information with SDM fibres}

\subsection{High-dimensional quantum key distribution over SDM fibres}

Many fundamental and applied tasks in quantum information benefit when $d$-dimensional ($d > 2$) quantum systems (qudits) are employed \cite{Kaszlikowski_2000, Araujo 2014, Martinez_2018}. One popular realisation for high-dimensional photonic quantum information processing is path encoding \cite{Rarity_1991, Neves_2005, Boyd_2005, Rossi_2009, Canas_2014, Obrien_2018, Martinez_2018}. A $d$-dimensional path-encoded qudit has the general form $|\psi\rangle =  \frac{1}{2}\sum_{d=0}^d e^{\phi_d} |d\rangle$, where $|d\rangle$ represents the $d_{th}$ path and $\phi_d$ is the relative phase in path $d$. A direct advantage gained in quantum communication when using qudits is increasing the transmission rate of QKD systems \cite{Cerf_2002}, due to the fact that we can encode $log_2 d$ bits onto an $d$-dimensional quantum state. This is a similar approach to what has always been done to increase the number of bits sent per symbol in classical communication systems. This technique is particularly useful, when it becomes too resource-intensive to simply increase the transmission rate by employing faster modulation and demodulation/detection opto-electronic devices. It then aims for a different approach, which takes advantage of extra dimensions  \cite{Proakis_2007}.

QKD's goal is to generate a shared secret string of bits (a key) between spatially separated parties (usually referred to as Alice and Bob) through the transmission of properly encoded single-photons \cite{Gisin_2002}. Current experiments in optical fibres can reach distances over 400 kms \cite{Boaron_2018}, and generate a few Mbit/s of final sifted key rate over metropolitan distances \cite{Patel_2014}. The security of QKD relies on the simple fact that an unknown quantum system cannot be faithfully cloned \cite{nocloning}, and this is explored in the well known BB84 protocol \cite{BB84}. Alice chooses randomly a quantum state to be transmitted and Bob does a random measurement on it, while recording the result. After many such rounds, Alice and Bob perform classical reconciliation and post-processing procedures to distill a shared secret key. Many factors affect the key generation rate, from the channel losses, the optical quality of the setup to the physical specifications of the detectors.

 \begin{figure}[t]
\centering
\includegraphics[width=0.9\textwidth]{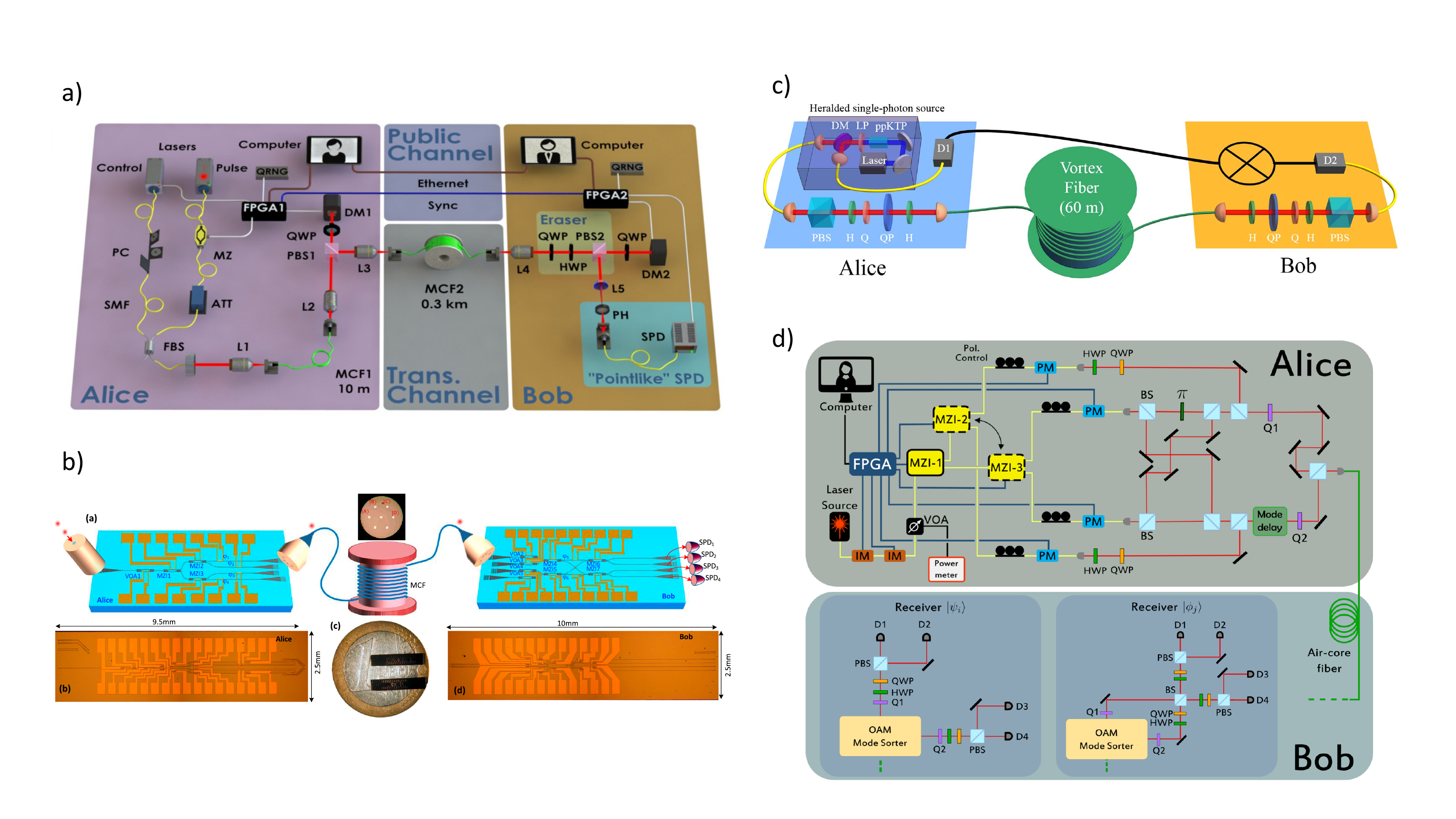}
\caption{Key experimental demonstrations of QKD with SDM fibre. a) 300-m-long HD-QKD session using path-encoding with a 4-core fibre. Reprinted with permission from \cite{Canas_2017}. b) HD-QKD session with path-encoded states based on integrated silicon photonic circuits using 4-cores in a 3-m-long fibre. Reprinted with permission from \cite{Ding_2017}. c) Proof-of-principle QKD demonstration using 2-dimensional OAM states through a 60 m vortex (ring-core) fibre. Reprinted with permission from \cite{Sit_2018}, [OSA]. d) HD-QKD demonstration using hybrid OAM/polarization states through 1.2 km air-core fibre. Reprinted with permission from \cite{Cozzolino_2018}.  \label{Fig2}}
\end{figure}

BB84-based proof-of-principle high-dimensional quantum key distribution (HD-QKD) experiments was already performed many years ago relying on the linear transverse momentum (LTM) of single-photons \cite{Steve_2006}. The two employed bases consisted of imaging and Fourier optical systems (by changing the appropriate lenses). The main limitation of this scheme is that the states had to be manually prepared and measured by changing the lenses and moving pinholes. An important next step occurred when spatial light modulators (SLMs) were used, capable of dynamically generating sets of parallel slits, which allowed the encoding of a high-dimensional qudit onto a single-photon propagating through these slits \cite{Glima_2009, Glima_2011}. This was then combined with synchronised FPGA (Field Programmable Gate Array) electronics to perform the first automated BB84 session in higher-dimensions using attenuated optical pulses \cite{Etcheverry_2013}. Here a QKD session using path-encoded 16-dimensional quantum states was realised by dynamically using the SLMs to prepare and measure the states, allowing 4 bits to be sent in each round. On the other hand, there has been considerable effort to implement HD-QKD using OAM encoding strategies through free-space \cite{Mafu_2013}, which was also later done with a fully automated setup \cite{Boyd_2015} and even over free-space intra-city channels \cite{Sit_2017}.

The major challenge of transmitting quantum systems encoded onto transverse optical modes through optical fibres remained. Some efforts had been done a few years ago using a 30-cm-long photonic crystal fibre \cite{Loffler_2011}, which are unfortunately not practical for long-distance propagation. The parallel development of SDM fibres came to change this paradigm, since spatial-mode-supporting fibres became widely available at a reasonable cost. Two experiments performed simultaneously kick-started this trend. In the first (Fig. \ref{Fig2}a \cite{Canas_2017}) a 300-m-long four-core fibre was used to perform a HD-QKD session using deformable mirrors as the phase modulators, yielding improvements over the previous efforts with SLMs \cite{Etcheverry_2013}. This result is the longest distance ever reported for the transmission of path-qudit states, showing that MCFs can be used for high-fidelity propagation over distances of practical interest. The other experiment also showed a successful HD-QKD session using a 4-core fibre, but with Alice and Bob's hardware fully implemented on integrated silicon photonics circuits \cite{Ding_2017} (Fig. \ref{Fig2}b). Thermal elements on the chip were employed to perform active modulation. Both experiments also carried out rigorous security analyses showing that much longer distances could be achieved while still being capable of positive secret key rates. It was also shown that MCFs can be used to deliver keys in parallel choosing separate sets of cores \cite{Bacco_2017}. More recently, ring-core-type fibres have also been employed for remote QKD sessions using OAM encoded qudits. The first one (Fig. \ref{Fig2}c) was actually done in a 2-dimensional space over a 60 m vortex fibre with no active state preparation \cite{Sit_2018}, while the second one (Fig. \ref{Fig2}d) employed hybrid polarisation/OAM states to perform an HD-QKD session using ququarts over 1.2 km-long air-core fibre \cite{Cozzolino_2018}.

\subsection{Entanglement distribution}

The successful distribution of photonic entanglement over long optical fibres is an important operational toolbox in quantum information. Many experiments have been performed using different degrees of freedom of a single-photon over optical fibres, mainly using polarisation \cite{Poppe_2004, Hubel_2007, Zhong_2010, Wengerowski_2019} and energy-time/time-bin \cite{Tittel_1998, Marcikic_2004, Cuevas_2013, Inagaki_2013}. For many years the distribution of spatial entanglement (i.e., entanglement between quantum systems encoded in terms of the transverse optical modes of light) over optical fibres has been out of reach. This has been mainly due to the fact that (i) single-mode fibres, by their very nature, do not support more than one spatial mode; and (ii) mode coupling scrambles the multi-mode spatial state during propagation. Nevertheless, significant progress has been made recently using both custom-made and standard commercially available optical fibres. In common, all experiments employ fibres that support only a few spatial modes, in order to minimize significant mode coupling.

The first experiment to be able to propagate spatially entangled photons employed a 30-cm long hollow-core photonic crystal fibre to transmit one of the photons of the pair, while successfully measuring a Bell inequality violation over the pair of qubits \cite{Loffler_2011}. This was an important first step demonstrating the feasibility of distributing spatially entangled states, but the use of specialty fibres hampers practical use. This was improved upon with standard polarisation maintaining fibres that were used as FMFs (by working with single-photons at 810 nm), and then two of the three possible linearly polarised modes in the fibre were used for each single-photon of the pair (LP$_{01}$ and the even LP$_{11}$ mode) \cite{Kang_2012}.

\begin{figure}[t]
\centering
\includegraphics[width=0.9\textwidth]{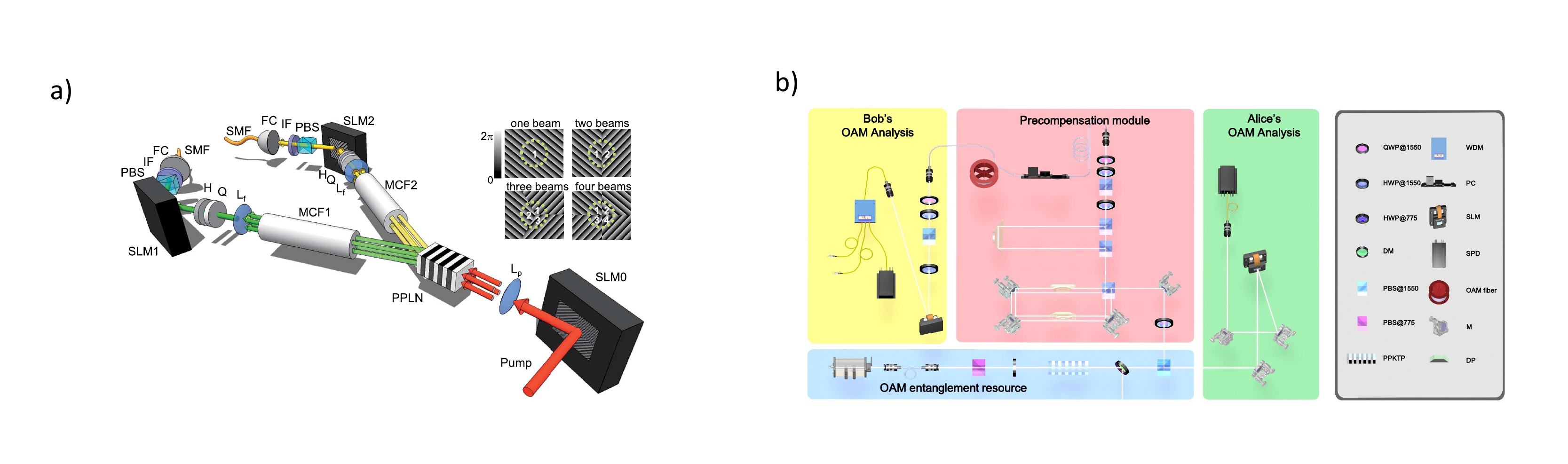}
\caption{High-dimensional distribution of spatially entangled states through optical fibres. a) Bell inequality test over 4-core fibres. Reprinted with permission from \cite{Lee_2019}. b) Distribution of 3-dimensional OAM entanglement over a 1 km long few-mode fibre. Reprinted with permission from \cite{Cao_2018}.  \label{Fig3new}}
\end{figure}

Nevertheless, one immediate advantage of resorting to spatial entanglement is the possibility of reaching higher-dimensional Hilbert spaces. When combined with fibre propagation, this allows the execution of more interesting high-dimensional quantum information tasks, where physical separation between parties who do not share line-of-sight is no longer a restriction. Going in this direction more recent experiments have taken direct advantage of the possibilities allowed by newly developed SDM hardware. In \cite{Lee_2017}, 4-dimensional spatial entanglement has been distributed over short-distances (30 cm) 4-core fibres. Entanglement was verified through quantum state tomography, and this was later expanded to a Bell inequality test \cite{Lee_2019} (Fig. \ref{Fig3new}a). Another experiment taking direct advantage of SDM fibres was \cite{Cui_2017}, where a 1 km-long FMF was used to test if polarisation and time-bin entanglement can be propagated through the different modes.

Very recent progress has also shown the propagation of OAM entangled quantum states over optical fibres. One of the photons of a 3-dimensional OAM entangled state was successfully propagated over a 1 km step-index fibre \cite{Cao_2018}, supporting up to 6 LP modes (Fig. \ref{Fig3new}b). The employed fibre required careful alignment of the input axis in order to properly excite and later decode the OAM modes. This experiment also featured compensation of modal dispersion, which is necessary as the fibre length increases, even more so for entangled states produced by down-conversion processes, due to the typical low temporal coherence involved. Another experiment started with a polarisation-entangled photon pair, where one of the photons gets further encoded with a superposition of OAM modes (vector vortex mode) to generate hybrid three-qubit entanglement \cite{Cozzolino_2019}. The hybrid-encoded photon is transmitted through a 5 m-long air-core fibre, showing that hybrid high-dimensional entanglement can be propagated over SDM fibres. Then, another experiment demonstrated hybrid OAM/polarisation entanglement with one of the photons propagated through 250 m of single-mode telecom fibre, which works as a FMF at the working wavelength in the experiment of 810 nm \cite{Liu_2019}. The main drawback here is the much greater attenuation ($>$ 2 dB/km) when compared to working with standard SDM FMF fibres at telecom wavelengths.

\begin{figure}[t]
\centering
\includegraphics[width=0.9\textwidth]{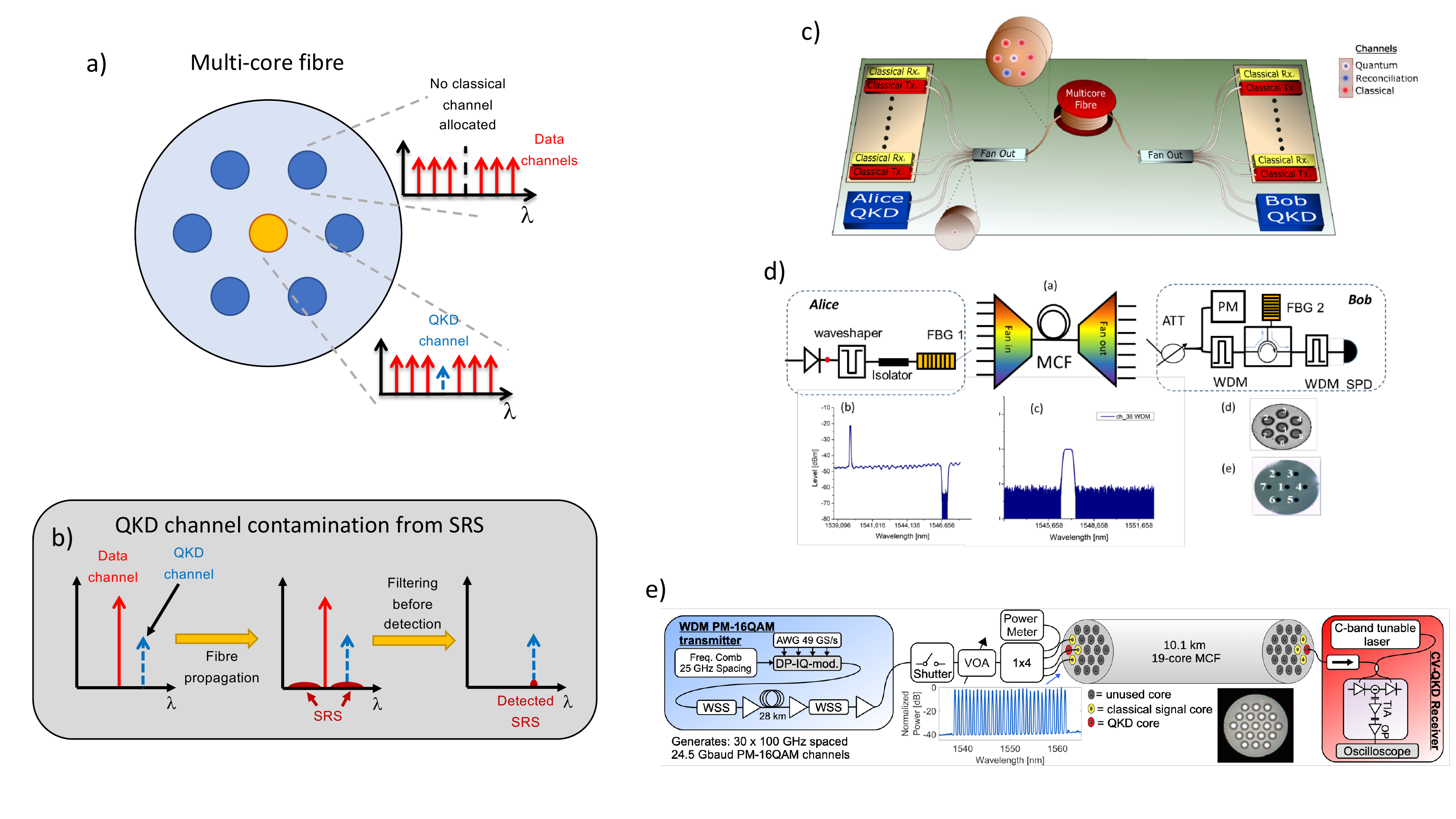}
\caption{General overview of co-existence of QKD and classical data channels over SDM fibres and key experiments. a) Wavelength channel allocation in a multi-core fibre. A wavelength is assigned to the QKD channel (in the centre core for example, shown in orange), and has to be kept free in the adjacent cores (shaded blue) to prevent in-band inter-core crosstalk \cite{Dynes_2016, Lin_2018, Hugues-Salas_2019, Cai_2019}. The other wavelengths in each core may be assigned to data channels. b) Spontaneous Raman scattering (SRS) generated when a classical channel is present in the same core can contaminate a QKD channel. Following propagation through the core of an MCF, SRS photons are produced from a classical channel (pump) over a broad wavelength range, eventually becoming strong noise to the QKD channel band. After the receiver's filtering stage, the SRS photons are detected as an increase in the dark counts of the QKD system, lowering the generated secret key rate. c) Experiment demonstrating co-existence of QKD and classical data over an MCF. Reprinted with permission from \cite{Dynes_2016} [OSA]. d) Experimental demonstration of SRS over all cores of NT- and TA-MCFs. Reprinted with permission from \cite{Lin_2019}. e) Co-existence experiment of CV-QKD and classical channels over an MCF. Reprinted with permission from \cite{Eriksson_2019}.  \label{Fig3}}
\end{figure}

\section{Integration with classical telecommunication optical networks}

Compatibility with optical networks is a major driving force for the widespread deployment of quantum communication \cite{Chapuran_2009, Patel_2012}. The next logical step is to ensure compatibility of quantum communication systems with the next-generation SDM optical networks. A first proof-of-principle experiment (over a short distance of 2 m) has shown that spatial multiplexing of a classical and a quantum channel is possible on a few-mode fibre \cite{Carpenter_2013}. Then a compatibility experiment between QKD and classical data co-existing in the same fibre using separate cores over a MCF was carried out \cite{Dynes_2016}. The centre core was reserved for the QKD channel, while the side cores were pairwise-filled with 10 Gbit/s data streams from opposite directions. A 7-core fibre was employed with a relatively high core-to-core distance ($47 \mu m$), such that inter-core crosstalk was rather low at -60 dB and -80 dB at the forward and backwards propagation directly. Nevertheless this crosstalk was still high enough to ensure that the quantum and classical channels could not share the same wavelengths on separate cores (Fig. \ref{Fig3}a).  If these limitations are taken into account, co-existence between quantum and classical signals in the same fibre is possible, as with single-mode fibres. A follow-up study characterised the difference in impact between trench-assisted and non-trench-assisted (TA- and NT- respectively) MCFs \cite{Lin_2018}, while having the same core-to-core distance in both fibres (41.1 $\mu m$). All cores were simultaneously loaded with high speed traffic (112 Gbit/s), at different wavelengths. NT-MCFs are interesting to be studied for compatibility, since they allow a higher core density due to the absence of lower refractive index trenches surrounding the cores. The results showed that compatibility is also possible, with a larger penalty imposed by the classical channels on the NT-MCF, as expected due to the higher cross-talk.

Recently, improved classical channel data rates has been achieved over a 7-core fibre while simultaneously allocating the center core to a QKD channel \cite{Hugues-Salas_2019}. Other works have also studied the use of a dedicated side-core for QKD while having neighbouring cores filled with classical data \cite{Cai_2019}, the inter-core crosstalk produced from classical channels on MCFs affecting continuous variable (CV) QKD systems \cite{Eriksson_2019} and performed detailed modelling on SDM-QKD integration \cite{Urena_2019}. Finally, a major increase in key generation rate has been achieved by sending out parallel keys over 37 cores of a MCF, while also propagating a 10Gbit/s data stream within each core, wavelength-multiplexed with the QKD channel \cite{Da Lio_2019}. One important limitation for co-existence is spontaneous Raman scattering (SRS), where photons from classical channels are inelastically scattered over a broad wavelength range \cite{Chapuran_2009}. Even if the cross-talk produced from in-band photons can be fully removed (with high-quality filtering for instance), SRS photons will eventually be scattered back to the QKD channel band over the fibre link (Fig. \ref{Fig3}b). Therefore, the overall system design needs to take this issue into account \cite{Patel_2012}. Recently, SRS has been demonstrated for the first time in each core of an MCF \cite{Lin_2019}, indicating as expected the same limitations when simultaneously propagating quantum and classical signals in the same core of an MCF. Table \ref{tab1} lists relevant quantum information experiments using SDM fibres.

\begin{table}[t]
\begin{center}
\begin{tabular}{c c c c c c c}
\hline
Year & Experiment & Fibre type & Distance & Clock rate & DOF & Reference\\

$2012$ & Entanglement distribution & FMF & 40 cm &** & Spatial &  \cite{Kang_2012} \\

$2013$ & Telecom integration & FMF & 2 m &** & ** &  \cite{Carpenter_2013} \\

$2016$ & QKD + telecom integration & 7-core MCF & 53 km &10 Gb/s & Time-bin &  \cite{Dynes_2016} \\

$2017$ & HD-QKD & 4-core MCF & 300 m &1 kHz & Path &  \cite{Canas_2017} \\

$2017$ & HD-QKD & 7-core MCF  (4 used) & 3 m &5 kHz & Path &  \cite{Ding_2017} \\

$2017$ & HD entanglement distribution & 4-core MCF & 30 cm &** & Path &  \cite{Lee_2017} \\

$2017$ & Entanglement distribution & FMF & 1 km &** & Polarisation/time-bin &  \cite{Cui_2017} \\

$2017$ & Parallel QKD & 7-core MCF (4 used) & 3 m &5 kHz & Path &  \cite{Bacco_2017} \\

$2018$ & QKD & Ring-core & 60 m & ** & OAM &  \cite{Sit_2018} \\

$2018$ & HD-QKD & Ring-core & 1.2 km & 50 kHz & OAM &  \cite{Cozzolino_2018} \\

$2018$ & Telecom integration & 7-core MCF & 2.5 km & ** & ** &  \cite{Lin_2018} \\

$2018$ & Parallel QKD & 37-core MCF & 7.9 km & 595 MHz & Time-bin &  \cite{Da Lio_2019} \\

$2018$ & HD entanglement distribution & FMF & 1 km & ** & OAM &  \cite{Cao_2018} \\

$2019$ & HD entanglement distribution & 4-core MCF & 30 cm & ** & Path &  \cite{Lee_2019} \\

$2019$ & QKD + telecom integration & 7-core MCF & 1 km & ** & ** &  \cite{Hugues-Salas_2019} \\

$2019$ & QKD + telecom integration & 7-core MCF & 30 km (attenuator) & 50 MHz & Time-bin &  \cite{Cai_2019} \\

$2019$ & Telecom integration & 19-core MCF & 10.1 km & ** & ** &  \cite{Eriksson_2019} \\

$2019$ & Telecom integration & 7-core MCF & 2.5 km & ** & ** &  \cite{Lin_2019} \\

$2019$ & HD entanglement distribution & Ring-core & 5 m & ** & Hybrid &  \cite{Eriksson_2019} \\

$2019$ & HD entanglement distribution & FMF & 250 m & ** & Hybrid &  \cite{Liu_2019} \\

$2019$ & MDI-QRNG & 4-core MCF & 45 cm & 2 MHz & Path &  \cite{Carine_2019} \\

\hline
\end{tabular}
\caption{Quantum information experiments using SDM fibres. Telecom integration refers to experiments aimed at characterising the impact of classical data channels on a quantum channel over SDM fibres. The clock rate refers to the repetition rates where active modulation was employed in the experiment. HD stands for high-dimensional, meaning that more than 2 dimensional systems were used, and DOF for degree-of-freesom. MDI-QRNG stands for measurement device-independent quantum random number generation. ** means ``not applicable'' or no explicit information available.}
\label{tab1}
\end{center}
\end{table}

\section{Outlook and open challenges}

There has been a rapid acceptance of SDM fibres and devices by the quantum information community with promising results reported so far. On the one hand the ability of these fibres to successfully manipulate and propagate high-dimensional quantum states over long distances has proven very fruitful for quantum information processing. On the other hand, their use for a number of QI protocols will ease the integration of both quantum and classical network systems based on the SDM optical fibre infrastructure.

Nevertheless, many challenges await that the community needs to handle to ensure that experiments based on SDM technology yield better results. Multi-core fibres have shown good promise for long-distance propagation and ease of integration with photonic circuits. Furthermore, high-quality devices can be built directly in the fibre itself \cite{Carine_2019}. Next it is important to see if they can support propagation distances for high-dimensional states that is comparable to single-mode fibres (i.e. hundreds of kms). They also need to be validated for states of even higher dimensionality, by employing a fibre with a higher core number, such as 7 or 19. Regarding telecom compatibility, it is important to verify more stringent limits such as noise produced from non-linear effects such as Raman scattering and four-wave mixing produced from classical channels in different wavelengths from different cores.

Although ring-core and even step-index fibres have also proved fruitful for OAM propagation, further experiments need to be carried out to verify their support for high-dimensions (only one experiment managed to go further than OAM qubits, and even then only OAM qutrits were employed \cite{Cao_2018}). This is worth pursuing since experiments in classical communications show that several modes can be simultaneously supported \cite{Wang_2018, Zhu_2018}, although MIMO detection was still required, so better mode isolation will be needed for QC experiments. By far most OAM based experiments have resorted to bulk-optics-based mode sorters, although recent results open up a new path towards integration for OAM \cite{Zeng_2018, Chen_2018}. FMFs also need to be studied in this regard, to verify their ability to coherently propagate superpositions of linearly polarised modes over long distances. The successful use of FMFs would open a direct path towards the transmission of much higher dimensional states, since they can be directly combined with MCFs. For instance, a fibre with 36 cores where each core supports the three lowest order LP modes, yielding a possible 108-dimensional space, has been demonstrated \cite{Sakaguchi_2016}.

An important challenge to be tackled comes from the modal dispersion. This problem is even more critical for entangled states due to their short coherence time. This has been tackled by pre-compensating the mode delay before transmission in the optical fibre \cite{Cozzolino_2018, Cao_2018}. Further efforts need to be carried out to demonstrate feasibility over much longer distances, as well as ability to cover compensation for a wide range of modes, aiming at high-dimensionality.

It is quite fortunate that the technological developments that are sustaining the growth of communication networks can be directly applied for quantum information technologies. This has been the case since the dawn of photonic quantum information, and will continue to be so. We have highlighted here the key developments in the nascent intersection of SDM and quantum information, with already many experiments having been able to directly benefit from SDM. We envisage that (i) integration between SDM networks and quantum information systems will be inevitable, and (ii) SDM offers the hardware to support efficient, high-fidelity, high-dimensional quantum information processing, which will be a major cornerstone of future developments in quantum technologies.

\subsection*{Acknowledgements}
G. X. acknowledges Ceniit Link\"{o}ping University and the Swedish Research Council (VR 2017-04470) for financial support. G. L. acknowledges the support of Fondecyt~1160400, and Millennium Institute for Research in Optics, MIRO.

\subsection*{Author contributions}
Both authors contributed equally to this manuscript.

\subsection*{Additional information}
Correspondence and requests for materials should be addressed to Guilherme B. Xavier or Gustavo Lima. The authors declare no competing financial interests.


\begin{references}

\bibitem{Feynman}
Feynman, R. P. Simulating physics with computers. \textit{Int. J. Theor. Phys.} \textbf{21}, 467 (1982).

\bibitem{Shor}
Shor, P. W. Polynomial-Time Algorithms for Prime Factorization and Discrete Logarithms on a Quantum Computer. \textit{SIAM J. Sci. Statist. Comput.} \textbf{26}, 1484 (1997).

\bibitem{nocloning}
Wootters, W. K., \& Zurek, W. H.  A single quantum cannot be cloned. \textit{Nature} \textbf{299}, 802 (1982).

\bibitem{BB84}
Bennett, C. H., \& Brassard. G. Quantum cryptography: Public key distribution and coin tossing. \textit{Proceedings of IEEE International Conference on Computers, Systems and Signal Processing} \textbf{175}, 8, New York (1984).

\bibitem{Gisin_2002}
Gisin, N., Ribordy, G., Tittel, W. \& Zbinden, H.
Quantum cryptography. \textit{Rev. Mod. Phys.} \textbf{74}, 145--195 (2002).

\bibitem{Lo_2014}
 Lo, H.-K., Curty, M. \and Tamaki, K. Secure quantum key distribution  \textit{Nat. Photon} \textbf{8}, 595 (2014).

\bibitem{Diamanti_2016}
Diamanti, E., Lo, H.-K., Qi, B. \& Yuan, Z. Practical challenges in quantum key distribution. \textit{npj Quantum Information} \textbf{2}, 16025 (2016).

\bibitem{Xu_2019}
Xu, F., Ma, X., Zhang, Q., Lo, H.-K. \& Pan, J.-W. Quantum cryptography with realistic devices. \textit{arXiv:1903.09051} (2019).

\bibitem{Gisin_2007}
Gisin, N. \& Thew, R. Quantum communication. \textit{Nat. Photon.} \textit{1}, 165 (2007).

\bibitem{Wehner_2018}
Wehner, S., Elkouss, D. \& Hanson, R. Quantum internet: A vision for the road ahead. \textit{Science} \textbf{362}, 303 (2018).

\bibitem{Agrawal}
Agrawal, G. P. Fiber-optic communication systems, third edition. Wiley, New York (2002).

\bibitem{WDM}
Brackett, C. A. Dense wavelength division multiplexing networks: principles and applications. \textit{IEEE J. Sel. Areas Commun.} \textbf{8}, 948 (1990).

\bibitem{Payne}
Mears, R. J., Reekie, L. Jauncey, I. M. \& Payne, D. N. Low-noise Erbium-doped fiber amplifier at 1.54$\mu$m. \textit{Electron. Lett.} \textbf{23}, 1026 (1987).

\bibitem{Desurvire}
Desurvire, E., Simpson, J. \& Becker, P. C. High-gain erbium-doped traveling-wave fiber amplifier. \textit{Opt. Lett.} \textbf{12}, 888 (1987).

\bibitem{Richardson_2013}
Richardson, D. J., Fini, J. M., \&  Nelson, L. E. Space-division multiplexing in optical fibres. \textit{Nat. Photon.} \textbf{7}, 354 (2013).

\bibitem{Wakayama_2019}
Wakayama, Y., Soma, D., Beppu, S., Sumita, S., Igarashi, K. \& Tsuritani, T. 266.1-Tbit/s Transmission Over 90.4-km 6-Mode Fiber With Inline Dual C+L-Band 6-Mode EDFA. \textit{IEEE J. Lightwave Technol.} \textbf{37}, 404 (2019).

\bibitem{Sciarrino_2018}
Flamini, F., Spagnolo, N. \& Sciarrino F. Photonic quantum information processing: a review. \textit{Rep. Prog. Phys.} \textbf{82}, 016001 (2018).

\bibitem{Erhard_2018}
Erhard, M., Fickler, R., Krenn, M. \& Zeilinger, A. Twisted photons: new quantum perspectives in high dimensions. \textit{Light Sci. \& Appl.} \textbf{7}, 17146 (2018).

\bibitem{Rarity_1991}
Rarity, J. G., \textit{et al}. Two-photon interference in a Mach-Zehnder interferometer. \textit{Phys. Rev. Lett.} \textbf{65}, 1348 (1991).

\bibitem{Neves_2005}
Neves, L., Lima, G., Aguirre G\'omez, Monken, C. H., Saavedra, C. \& P\'adua S. Generation of entangled states of qudits using twin photons. \textit{Phys. Rev. Lett.} \textbf{94}, 100501 (2005).

\bibitem{Boyd_2005}
O'Sullivan-Hale, M. N., Ali Khan, I., Boyd, R. W. \& Howell, J. C. Pixel entanglement: experimental realization of optically entangled d= 3 and d= 6 qudits. \textit{Phys. Rev. Lett.} \textbf{94}, 220501 (2005).

\bibitem{Groblacher_2006}
Gr\"{o}blacher, S., Jennewein, T., Vaziri, A., Weihs, G. \& Zeilinger, A.
Experimental quantum cryptography with qutrits.
\textit{New J. Phys.} \textbf{8}, 75 (2006).

\bibitem{Rossi_2009}
Rossi, A., Vallone, G., Chiuri, A., De Martini F. \& Mataloni, P. Multipath entanglement of two photons. \textit{Phys. Rev. Lett.} \textbf{102}, 153902 (2009).

\bibitem{Aguilar_2018}
Aguilar, E. A., \textit{et al}. Certifying an irreducible 1024-dimensional photonic state using refined dimension witnesses. \textit{Phys. Rev. Lett.} \textbf{120}, 230503 (2018).

\bibitem{Duan_2001}
Duan, L.-M., Lukin, M. D., Cirac, J. I. \& Zoller, P. Long-distance quantum communication with atomic ensembles and linear optics. \textit{Nature} \textbf{414}, 413 (2001).

\bibitem{Politi_2008}
Politi, A., Cryan, M. J., Rarity, J. G., Yu, S. \& O'Brien J. L. Silica-on-Silicon Waveguide Quantum Circuits. \textit{Science} \textbf{320}, 646 (2008).

\bibitem{Obrien_2018}
Wang, J., \textit{et al}. Multidimensional quantum entanglement with large-scale integrated optics. \textit{Science} \textbf{360}, 285 (2018).

\bibitem{Chen_2018}
Chen, Y., \textit{et al}. Mapping Twisted Light into and out of a Photonic Chip. \textit{Phys. Rev. Lett.} \textbf{121}, 233602 (2018).


\bibitem{Choi_2012}
Choi, Y., \textit{et al}. Scanner-Free and Wide-Field Endoscopic Imaging by Using a Single Multimode Optical Fiber. \textit{Phys. Rev. Lett.} \textbf{109}, 203901 (2012).

\bibitem{Iano_1979}
Iano, S., Sato, T., Sentsui, S., Kuroha T. \& Nishimura, Y. Multicore optical fiber. \textit{In Optical Fiber Communication Conference (OFC) 1979, OSA Technical Digest (Optical Society of America, 1979)} paper WB1 (OSA, 1979).

\bibitem{Saitoh_2016}
Saitoh, K. \& Matsuo, S. Multicore Fiber Technology. \textit{IEEE J. Lightwave Technol.} \textbf{34}, 55 (2016).

\bibitem{MIMO}
Foschini, G. J. Layered space-time architecture for wireless communication in a fading environment when using multi-element antennas. \textit{Bell Labs Tech. J.} \textbf{1}, 41 (1996).

\bibitem{Saitoh_2013}
Saitoh, K. \& Matsuo, S. Multicore fibers for large capacity transmission. \textit{Nanophotonics} \textbf{2}, 441 (2013).

\bibitem{Sakaguchi_2013}
Sakaguchi, J., \textit{et al}.19-core MCF transmission system using EDFA with shared core pumping coupled via free-space optics. \textit{Opt. Express} \textbf{22}, 90 (2013).


\bibitem{Berdague_1982}
Berdagu\'e, S. \& Facq, P. Mode division multiplexing in optical fibers. \textit{Appl. Opt.} \textbf{21}, 1950 (1982).




\bibitem{Sillard_2014}
Sillard, P., Bigot-Astruc, M. \& Molin, D. Few-Mode Fibers for Mode-Division-Multiplexed Systems. \textit{IEEE J. Lightwave Technol.} \textbf{32}, 2824 (2014).

\bibitem{Xia_2012}
Xia, C., \textit{et al}. Hole-Assisted Few-Mode Multicore Fiber for High-Density Space-Division Multiplexing. \textit{IEEE Photonic. Tech. Lett.} \textbf{24}, 1914 (2012).

\bibitem{van_Uden_2014}
van Uden, R. G. H., \textit{et al}. Ultra-high-density spatial division multiplexing with a few-mode multicore fibre. \textit{Nat. Photon.} \textbf{8}, 865 (2014).

\bibitem{Gibson_2004}
Gibson, G., \textit{et al}. Free-space information transfer using light beams carrying orbital angular momentum. \textit{Opt. Express} \textbf{12}, 5448 (2004).


\bibitem{Allen_1992}
Allen, L., Beijersbergen, M. W., Spreeuw, R. J. C. \& Woerdman, J. P. Orbital angular momentum of light and the transformation of Laguerre-Gaussian laser modes. \textit{Phys. Rev. A} \textbf{45}, 8185 (1992).

\bibitem{Wang_2012}
Wang, J., \textit{et al}. Terabit free-space data transmission employing orbital angular momentum multiplexing. \textit{Nat. Photon.} \textbf{6}, 488 (2012).

\bibitem{Bozinovic_2013}
Bozinovic, N. \textit{et al}. Terabit-Scale Orbital Angular Momentum Mode Division Multiplexing in Fibers. \textit{Science} \textbf{340}, 1545 (2013).

\bibitem{Brunet_2015}
Brunet, C., Ung, B., Wang, L., Messaddeq, Y., LaRochelle, S. \& Rusch, L. A. Design of a family of ring-core fibers for OAM transmission studies. \textit{Opt. Express} \textbf{23}, 10553 (2015).

\bibitem{Gregg_2015}
Gregg, P., Kristensen, P. \& Ramachandran, S. Conservation of orbital angular momentum in air-core optical fibers. \textit{Optica} \textbf{2}, 267 (2015).

\bibitem{Thomson_2012}
Thomson, R. R., Harris, R. J., Birks, T. A., Brown, G., Allington-Smith, J. \& Bland-Hawthorn, J. Ultrafast laser inscription of a 121-waveguide fan-out for astrophotonics. \textit{Opt. Lett.} \textbf{37}, 2331 (2012).

\bibitem{Watanabe_2012}
Watanabe, K., Saito, T., Imamura, K. \& Shiino, M. Development of fiber bundle type fan-out for multicore fiber. \textit{In IEEE 17th Opto-Electronics and Communications Conference} (2012).

\bibitem{Tottori_2012}
Tottori, Y., Kobayashi, T. \& Watanabe, M. Low Loss Optical Connection Module for Seven-Core Multicore Fiber and
Seven Single-Mode Fibers. \textit{IEEE Photon. Tech. Lett.} \textbf{24}, 1926 (2012).

\bibitem{Birks_2015}
Birks, T. A., Gris-S\'anchez, I., Yerolatsitis, S., Leon-Saval, S. G. \& R. R. Thomson. The photonic lantern. \textit{Adv. Opt. Photonics} \textbf{7}, 107 (2015).

\bibitem{Yerolatsitis_2014}
Yerolatsitis, S., Gris-S\'anchez, I. \& Birks, T. A. Adiabatically-tapered fiber mode multiplexers. \textit{Opt. Express} \textbf{22}, 608 (2014).

\bibitem{Riesen_2014}
Riesen, R. R., Gross, S., Love, J. D. \& Withford, M. J. Femtosecond direct-written integrated mode
couplers. \textit{Opt. Express} \textbf{22}, 29855 (2014).

\bibitem{Berkhout_2010}
Berkhout, G. C. G., Lavery, M. P. J., Courtial, J., Beijersbergen, M. W. \& Padgett, M. J. Efficient Sorting of Orbital Angular Momentum States of Light. \textit{Phys. Rev. Lett.} \textbf{105}, 153601 (2010).

\bibitem{Zeng_2018}
Zeng, X., \textit{et al}. All-fiber orbital angular momentum mode multiplexer based on a mode-selective photonic lantern and a mode polarization controller. \textit{Opt. Lett.} \textbf{43}, 4779 (2018).

\bibitem{Riesen_2017}
Riesen, N., Gross, S., Love, J. D., Sasaki, Y. \& Withford, M. J. Monolithic mode-selective few-mode multicore fiber multiplexers. \textit{Sci. Rep.} \textbf {7}, 6971 (2017).

\bibitem{Kaszlikowski_2000}
Kaszlikowski, D., Gnaci\'nski, P., \.Zukowski, M. Miklaszewski, W. \& Zeilinger A. Violations of Local Realism by Two Entangled \textit{N}-Dimensional Systems Are Stronger than for Two Qubits. \textit{Phys. Rev. Lett.} \textbf{85}, 4418 (2000).

\bibitem{Araujo 2014} Ara\'ujo, M., Costa, F. \& Brukner, C. Computational Advantage from Quantum-Controlled Ordering of Gates. \textit{Phys. Rev. Lett.} \textbf{113}, 250402 (2014).

\bibitem{Martinez_2018}
Martinez, D., Tavakoli, A., Casanova, M., Ca\~{n}as, G. \& Lima, G. High-dimensional quantum communication complexity beyond strategies based on Bell's theorem. \textit{Phys. Rev. Lett.} \textbf{121}, 150504 (2018).

\bibitem{Canas_2014}
Ca\~nas, \textit{et al}. Applying the simplest Kochen-Specker set for quantum information processing. \textit{Phys. Rev. Lett.} \textbf{113}, 090404 (2014).

\bibitem{Cerf_2002}
Cerf, N. J., Bourennane, M., Karlsson, A. \& Gisin, N. Security of quantum key distribution using $d$-level systems \textit{Phys. Rev. Lett.} \textbf{88}, 127902 (2002).

\bibitem{Proakis_2007}
Proakis, J. G. \& Salehi, M.
\textit{Digital Communications 5th Edition} (McGraw-Hill 2007).

\bibitem{Boaron_2018}
Boaron, A., \textit{et al}. Secure Quantum Key Distribution over 421 km of Optical Fiber.\textit{Phys. Rev. Lett.} \textbf{121}, 190502 (2018).

\bibitem{Patel_2014}
Patel, K. A., \textit{et al}. Quantum key distribution for 10 Gb/s dense wavelength division multiplexing networks. \textit{Appl. Phys. Lett.} \textbf{104}, 051123 (2014).


\bibitem{Steve_2006}
Walborn, S. P., Lemelle, D. S., Almeida, M. P. \& Souto Ribeiro, P. H.
Quantum key distribution with higher-order alphabets using spatially encoded qudits.
\textit{Phys. Rev. Lett.} \textbf{96}, 090501 (2006).

\bibitem{Glima_2009}
Lima, G., Vargas, A., Neves, L., Guzm\'an, R. \& Saavedra, C.
Manipulating spatial qudit states with programmable optical devices.
\textit{Opt. Express}, \textbf{17}, 10688 (2009).

\bibitem{Glima_2011}
Lima, G., \textit{et al}. Experimental quantum tomography of photonic qudits via mutually unbiased basis. \textit{Opt. Express}, \textbf{19}, 3542 (2011).

\bibitem{Etcheverry_2013}
Etcheverry, S. \textit{et al}.
 Quantum key distribution session with 16-dimensional photonic states
\textit{Sci. Rep.} \textbf{3}, 2316 (2013).

\bibitem{Mafu_2013}
Mafu, M. \textit{et al}. Higher-dimensional orbital-angular-momentum-based quantum key distribution with mutually unbiased bases. \textit{Phys. Rev. A} \textbf{88}, 032305 (2013).

\bibitem{Boyd_2015}
Mirhosseini, M. \textit{et al}. High-dimensional quantum cryptography with twisted light. \textit{New. J. Phys.} \textbf{17}, 033033 (2015).

\bibitem{Sit_2017}
Sit, A., \textit{et al}. High-dimensional intracity quantum cryptography with structured photons. \textit{Optica} \textbf{4}, 1006 (2017).

\bibitem{Loffler_2011}
L\"{o}ffler, W., Euser, T. G., Eliel, E. R., Scharrer, M., Russell, P. St. J. \& Woerdman, J. P. Fiber Transport of Spatially Entangled Photons. \textit{Phys. Rev. Lett.} \textbf{106}, 240505 (2011).

\bibitem{Canas_2017}
Can\~as, G. \textit{et al}. High-dimensional decoy-state quantum key distribution over multicore telecommunication fibers. \textit{Phys. Rev. A} \textbf{96}, 022317 (2017).

\bibitem{Ding_2017}
Ding, Y. \textit{et al}. High-dimensional quantum key distribution based on multicore fiber using silicon photonic integrated circuits. \textit{npj Quantum Information} \textbf{3}, 25 (2017).

\bibitem{Bacco_2017}
Bacco, D., Ding, Y., Dalgaard, K., Rottwitt, K. \& Oxenl\o we. Space division multiplexing chip-to-chip quantum key distribution. \textit{Sci. Rep.} \textbf{7}, 12459 (2017).

\bibitem{Sit_2018}
Sit, A. \textit{et al}. Quantum cryptography with structured photons through a vortex fiber. \textit{Opt. Lett.} \textbf{43}, 4108 (2018).

\bibitem{Cozzolino_2018}
Cozzolino, D. \textit{et al}. Fiber based high-dimensional quantum communication with twisted photons. \textit{arXiv:1803.10138v1} (2018).

\bibitem{Poppe_2004}
Poppe, A., \textit{et al}. Practical quantum key distribution with polarization entangled photons. \textit{Opt. Express} \textbf{12}, 3865 (2004).

\bibitem{Hubel_2007}
H\"{u}bel, H., \textit{et al}. High-fidelity transmission of polarization encoded qubits from an entangled source over 100 km of fiber. \textit{Opt. Express} \textbf{15}, 7853 (2007).

\bibitem{Zhong_2010}
Zhong, T., Hu, X., Wong, F. N. C., Berggren, K. K., Roberts, T. D. \& Battle, P. High-quality fiber-optic polarization entanglement distribution at $1.3 ~\mu$m  telecom wavelength. \textit{Opt. Lett.} \textbf{35}, 1392 (2010).

\bibitem{Wengerowski_2019}
Wangerowski, S., \textit{et al}. Entanglement distribution over a 96-km-long submarine optical fiber. \textit{Proc. Natl. Acad. Sci. USA} \textbf{116}, 6684 (2019).

\bibitem{Tittel_1998}
 Tittel, W., Brendel, J., Zbinden, H. \& Gisin, N.
 Violation of Bell inequalities by photons more than 10 km apart.
 \textit{Phys. Rev. Lett.} \textbf{81}, 3563--3566 (1998).

\bibitem{Marcikic_2004}
 Marcikic, I., de Riedmatten, H., Tittel, W., Zbinden, H., Legr\'{e}, M. \& Gisin, N.
 Distribution of time-bin entangled qubits over 50 km of optical fiber.
 \textit{Phys. Rev. Lett.} \textbf{93}, 180502 (2004).


\bibitem{Cuevas_2013}
Cuevas, A., \textit{et al}. Long-distance distribution of genuine energy-time entanglement. \textit{Nat. Commun.} \textit{4}, 2871 (2013).

\bibitem{Inagaki_2013}
Inagaki, T., Matsuda, N., Tadanaga, O., Asobe, M. \& Takesue, H. Entanglement distribution over 300 km of fiber. \textit{Opt. Express} \textbf{21}, 23241 (2013).

\bibitem{Kang_2012}
Kang, Y., Ko, J., Lee, S. M., Choi, S.-K., Kim, B. Y. \& Park, H. S. Measurement of the Entanglement between Photonic Spatial Modes in Optical Fibres. \textit{Phys. Rev. Lett.} \textbf{109}, 020502 (2012).

\bibitem{Lee_2017}
Lee, H. J., Choi, S.-K. \& Park, H. S. Experimental Demonstration of Four-Dimensional Photonic Spatial Entanglement between Multi-core Optical Fibres. \textit{Sci. Rep. } \textbf{7}, 4302 (2017).

\bibitem{Lee_2019}
Lee, H. J. \& Park, H. S. Generation and measurement of arbitrary four-dimensional spatial entanglement between photons in multicore fibers. \textit{Photon. Res.} \textbf{7}, 19 (2019).

\bibitem{Cui_2017}
Cui, L., Su, J., Li, X. \& Ou, Z. Y. Distribution of entangled photon pairs over few-mode fibers. \textit{Sci. Rep.} \textbf{7}, 14954 (2017).

\bibitem{Cao_2018}
Cao, H., \textit{et al}. Distribution of high-dimensional orbital angular momentum entanglement at telecom wavelength over 1km OAM fiber. \textit{arXiv:1811.12195v1} (2018).

\bibitem{Cozzolino_2019}
Cozzolino, D., \textit{et al}. Air-core fiber distribution of hybrid vector vortex-polarization entangled states. \textit{arXiv:1903.03452v1} (2019).

\bibitem{Liu_2019}
Liu, J., Nape, I., Wang, Q., Vall\'es, A., Wang, J. \& Forbes, A. Multi-dimensional entanglement transport through single-mode fibre. \textit{arXiv:1904.03114v1} (2019).

\bibitem{Chapuran_2009}
Chapuran, T. E., \textit{et al}. Optical networking for quantum key distribution and quantum communications. \textit{New J. Phys.} \textbf{11}, 105001 (2009).

\bibitem{Patel_2012}
Patel, K. A., \textit{et al}. Coexistence of High-Bit-Rate Quantum Key Distribution and Data on Optical Fiber. \textit{Phys. Rev. X} \textbf{2}, 041010 (2012).

\bibitem{Carpenter_2013}
Carpenter, J., \textit{et al}. Mode multiplexed single-photon and classical channels in a few-mode fiber. \textit{Opt. Express} \textbf{21}, 28794 (2013).

\bibitem{Dynes_2016}
Dynes, J. F. \textit{et al}. Quantum key distribution over multicore fiber. \textit{Opt. Express} \textbf{24}, 8081 (2016).

\bibitem{Lin_2018}
Lin, R. \textit{et al}. Telecom Compatibility Validation of Quantum Key Distribution Co-existing with 112 Gbps/$\lambda$/core Data Transmission in Non-Trench and Trench-Assistant Multicore Fibers. 2018 European Conference on Optical Communications (ECOC), paper We1A.3.

\bibitem{Hugues-Salas_2019}
Hugues-Salas, E., Wang, R., Kanellos, G. T., Nejabati, R. \& Simeonidou, D. Co-existence of 9.6 Tb/s Classical Channels and a Quantum Key Distribution (QKD) Channel over a 7-core Multicore Optical Fibre. \textit{In 2018 IEEE British and Irish Conference on Optics and Photonics (BICOP)}. (IEEE, 2019).

\bibitem{Cai_2019}
Cai, C., Sun, Y., Zhang, Y., Zhang, P., Niu, J. \& Ji, Y. Experimental wavelength-space division multiplexing of quantum key distribution with classical optical communication over multicore fiber. \textit{Opt. Express} \textbf{27}, 5125 (2019).

\bibitem{Eriksson_2019}
Eriksson, T. A., \textit{et al}. Inter-Core Crosstalk Impact of Classical Channels on CV-QKD in Multicore Fiber Transmission. \textit{In Optical Fiber Communication Conference (OFC) 2019, OSA Technical Digest (Optical Society of America, 2019)} paper Th1J.1 (OSA, 2019).

\bibitem{Urena_2019}
Ure\~{n}a, M., Gasulla, I., Fraile, F. J. \& Capmany, J. Modeling optical fiber space division
multiplexed quantum key distribution systems. \textit{Opt. Express} \textbf{27}, 7047 (2019).

\bibitem{Da Lio_2019}
Da Lio, B., \textit{et al}. Record-High Secret Key Rate for Joint Classical and Quantum Transmission Over a 37-Core Fiber. \textit{In 2018 IEEE Photonics Conference (IPC)}. (IEEE 2018).

\bibitem{Lin_2019}
Lin, R., \textit{et al}. Spontaneous Raman Scattering Effects in Multicore Fibers: Impact on Coexistence of Quantum and Classical Channels. \textit{In Optical Fiber Communication Conference (OFC) 2019, OSA Technical Digest (Optical Society of America, 2019)} paper M4C.2 (OSA, 2019).

\bibitem{Carine_2019}
Cari\~ne, J. \textit{et al}. \textit{in preparation} (2019).

\bibitem{Wang_2018}
Wang, A., Zhu, L., Wang, L., Ai, J., Chen, S. \& Wang, J. Directly using 8.8-km conventional multi-mode fiber for 6-mode orbital angular momentum multiplexing transmission. \textit{Opt. Express} \textbf{26}, 10038 (2018).

\bibitem{Zhu_2018}
Zhu, G., \textit{et al}. Scalable mode division multiplexed transmission over a 10-km ring-core fiber using high-order orbital angular momentum modes. \textit{Opt. Express} \textbf{26}, 594 (2018).

\bibitem{Sakaguchi_2016}
Sakaguchi, J., \textit{et al}. Large Spatial Channel (36-Core × 3 mode) Heterogeneous Few-Mode Multicore Fiber. \textit{IEEE J. Lightwave Tech.} \textbf{34}, 93 (2016).










\end{references}
\end{document}